\documentclass[useAMS,usenatbib]{mn2e}
\usepackage{graphicx}

\title[A new population of extended, luminous star clusters in the halo of M31]
{A new population of extended, luminous star clusters in the halo of M31}
\author[Huxor et al.]{A. P. Huxor$^{1}$, N. R. Tanvir$^{1}$, M. J. Irwin$^{2}$, R. Ibata$^{3}$,
\newauthor J. L. Collett$^{1}$, A. M. N. Ferguson$^{4}$, T. Bridges$^{5}$, G. F. Lewis$^{6}$\\
$^{1}$Centre for Astrophysics Research, University of Hertfordshire, Hatfield AL10 9AB; 
ahuxor@star.herts.ac.uk \\
$^{2}$Institute of Astronomy, Madingley Road, Cambridge, CB3 0HA \\
$^{3}$Observatoire de Strasbourg, 11, rue de l'Universite, F-67000, Strasbourg, France \\
$^{4}$Institute for Astronomy, University of Edinburgh, Royal Observatory, Blackford Hill, Edinburgh, EH9 3HJ\\
$^{5}$Department of Physics, Queen's University, Kingston, Ontario, Canada K7M 3N6 \\
$^{6}$Institute of Astronomy, School of Physics, A29, University of Sydney, NSW 2006, Australia} 

\begin{document}

\pagerange{\pageref{firstpage}--\pageref{lastpage}} \pubyear{2004}

\maketitle

\label{firstpage}

\begin{abstract}
We present three new clusters discovered in the halo of M31 which, although having globular-like 
colours and luminosities,
have unusually large half-light radii, $\sim$ 30 pc. They lie at projected galactocentric distances 
of $\approx$ 15 to $\approx$ 35 kpc. These objects begin to fill the gap in parameter space between
globular clusters and dwarf spheroidals, and are unlike any clusters found in the Milky Way, or elsewhere to date.
Colour-magnitude diagrams, integrated photometric properties and derived King profile fit parameters 
are given, and we discuss possible origins of these clusters
and their relationships to other populations.

\end{abstract}


\section{Introduction}

Over the past several years we have undertaken a major survey of M31, which has 
revealed a wealth of unexpected substructure \citep{Fergusonetal02}, 
the most prominent being a giant stream of stars near the minor axis \citep{Ibataetal01}. 
As part of this study of M31, we have searched for globular clusters (GCs) in a large part of the halo.
GC systems have been shown to be valuable tools for the study of the evolution of their
host galaxies, acting as chemical and dynamical probes \citep{Westetal04}. 
Specifically, most GCs are believed to be old objects, and thus 
provide clues to the earliest epochs of galaxy formation history.

The three clusters presented here were discovered during this search for classical GCs, 
whose results are more fully documented in another
 paper (Huxor et al. in prep). 
While undertaking aperture photometry of one of the GCs found, a diffuse cluster was 
serendipitously discovered nearby in the same field (cluster 1, hereafter C1, in figure 
\ref{Fi:clusters}). 
The object was not classed as a single object by the INT-WFS 
(Isaac Newton Telescope Wide Field Survey)
pipeline \citep{IrwinLewis01}, and hence,
had been missed by the semi-automated techniques that we employed
in the main GC survey. It is very distinctive in being significantly more
extended than typical GCs: aperture photometry giving a half-light
radius ($R_{h}$) of 35 pc, assuming a distance to M31 of $\sim$780 kpc \citep{McConnachieetal05}. 
We were therefore motivated to undertake a visual survey of
all the INT-WFS images for the M31 region, which resulted in the discovery of 
two further objects  with similar properties. 
In this paper we present the three clusters, which are unlike any found around the Milky Way, M31 or 
indeed elsewhere to date, in that although many MW GCs have half-light radii approaching 30 pc, they are 
faint. These new clusters, however, are of average GC lumonosity.

\section{The New Clusters}

\begin{figure*}
 \includegraphics[angle=0,width=100mm]{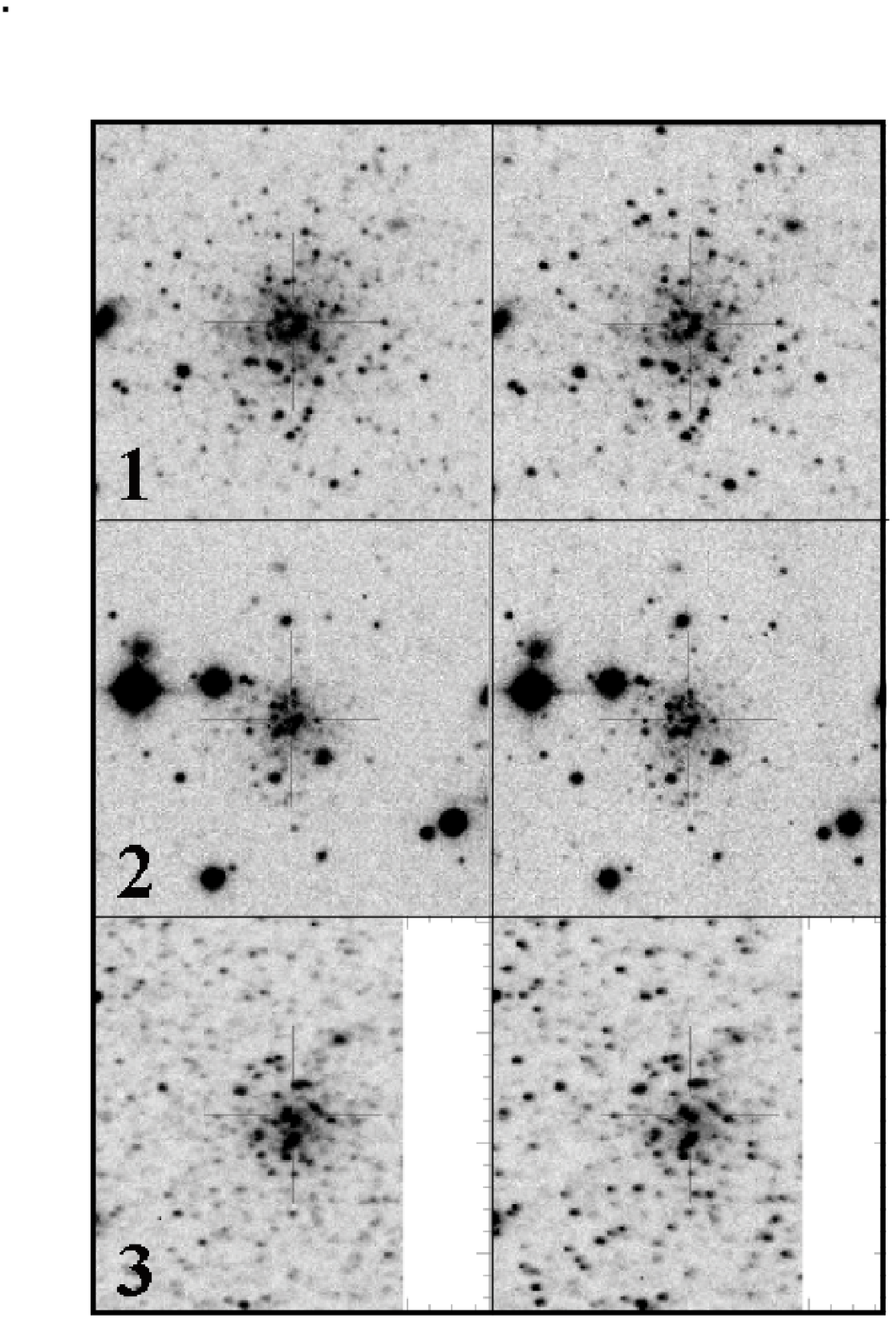}
 \vspace{2pt}
 \caption{V and i band images of the new luminous 'extended' clusters, from 
 the INT Wide Field Survey images.  Each image is 1 arcmin $\times$ 1 arcmin, 
 with North up and East to the left. Cluster 3 is a partial image as it lies 
 on the edge of an INT-WFS field.}\label{Fi:clusters}
\end{figure*}

\begin{figure}
 \includegraphics[angle=0,width=85mm]{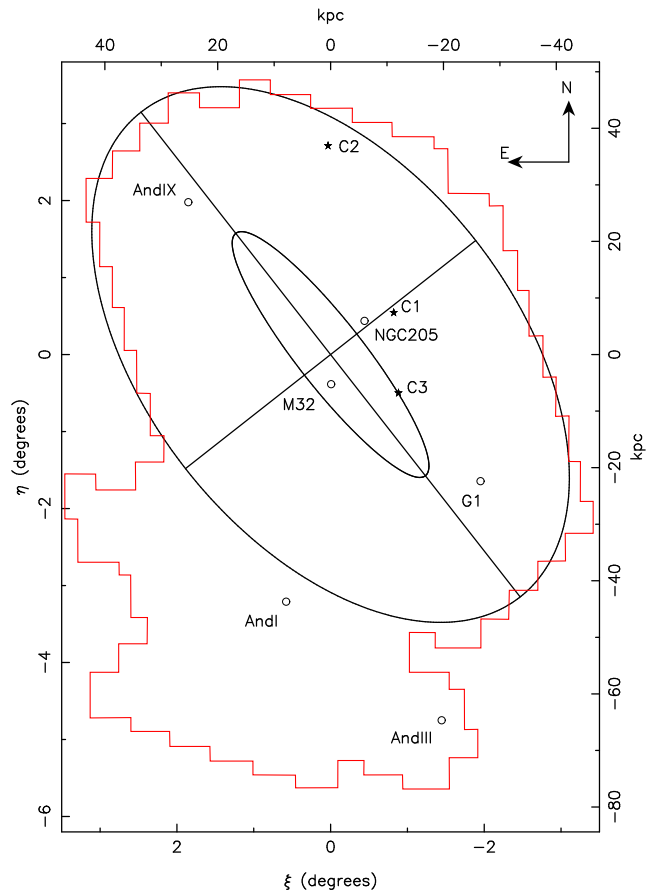}
 \vspace{2pt}
 \caption{The location of the new extended globular clusters 
 (stars) in relation to major landmarks of the M31 system (open circles), 
 and the survey area (dotted outline, red in electronic version). 
 The ellipses represent a 2 degree 
 radius disk aligned and tilted to the inclination of M31 and an oblate halo 
 of axial ratio 0.6 aligned along the major axis. 
 The kpc scales correspond to a distance to M31 of 780 kpc.}\label{Fi:layout}
\end{figure}

The fields visually investigated include the whole INT-WFS M31 survey, 
an area far into the halo, and an additional region 
south along the Andromeda Stream, and towards M33, making a total area of more 
than 40 square degrees (see figure \ref{Fi:layout}). 
The survey consists of V and Gunn i band images
with exposures of between 800-1000 seconds, reaching 
(average 5 $\sigma$) limiting magnitudes of i = 23.5 and V =24.5, and taken 
in average seeing of 1.2 arcsec. These images were processed by the INT WFS 
pipeline provided by the Cambridge Astronomical Survey Unit, which includes 
tools for astrometry, photometry and object description and classification 
\citep{IrwinLewis01}.  

The location of the clusters (star icons), in relation to M31 is shown in 
figure \ref{Fi:layout}, while postage stamp images (1 arcmin $\times$ 1 arcmin)
of the clusters are shown in figure \ref{Fi:clusters}.
The clusters appear, in projection, to cover a range of environments, 
from  the far halo (C2), to near the disk (C3; though we note that if C3 
is a disk object, it lies at a very large radius of 27 kpc). 

The positions of the three extended globular clusters were cross-checked 
against existing catalogues starting with the GC lists of Barmby 
(http://cfa-www.harvard.edu/$\sim$pbarmby/m31gc.html; \citealt{Barmbyetal00}) 
and the newly published Revised Bologna Catalog 
(http://www.bo.astro.it/M31/; \citealt{Galletietal04a}).  In addition, other 
databases such as Vizier \citep{Ochsenbeinetal00} and NED 
(http://nedwww.ipac.caltech.edu/index.html) 
were consulted for any prior identifications.  C1, the richest cluster, has 
previously been catalogued from a large scale survey as a background low surface 
brightness galaxy, object N04-1 of
\citet{ONeilBothunCornell97}.  However, the INT-WFS survey image clearly 
shows it to be a resolved star cluster.

CMDs of the clusters (see figures \ref{Fi:cmd1}, \ref{Fi:cmd2} and 
\ref{Fi:cmd3}) were made from the INT-WFS images by performing
DAOPHOT (IRAF implementation) profile-fitting photometry on the regions
of the images containing the clusters.
This level of crowding is an ideal application of profile-fitting photometry,
and the large number of bright stars on the images allows a good PSF model to be
constructed in each case.
The resulting
CMDs are fully consistent with metal-poor, old stellar 
populations at the distance of M31.  There is no evidence for a population
of young main sequence stars at these locations. The integrated $(V-I)_{0}$ colours are
also consistent with low metallicity (\citealt{Worthey94}) if indeed they are an old population.

\begin{table*}
 \centering
  \caption{
  Properties of the Extended Clusters. $R_{h}$ is the half-light 
  radius calculated assuming a distance to M31 of 780 kpc, which is also used
   the determination of absolute magnitude.  It is not easy 
  to determine accurate uncertainties on the values of $R_{h}$, given the sparseness 
  of the clusters, and the necessity to account for foreground and background 
  contamination, however we estimate the accuracy to be better than 
  $\approx$20\%. Extinction corrected magnitudes use the \citet{Schlegeletal98} maps.
  }
  \begin{tabular}{@{}llllllll@{}}
  \hline
   ID & RA (J2000) & Dec (J2000) & V & $V_{0}$ &($V-I)_{0}$ & $R_{h}$ & $M_{V}$ \\
      &  h     m     s   &    $\degr$     $\arcmin$     $\arcsec$  &  & &   & (pc) & \\

 \hline
 M31WFS-C1   & 00 38 19.5 & +41 47 15 & 17.6 & 17.4 & 0.88 & 34 & -7.1 \\
 M31WFS-C2   & 00 42 55.0 & +43 57 28 & 17.1 & 16.8 & 0.93 & 26 & -7.7 \\
 M31WFS-C3   & 00 38 04.6 & +40 44 39 & 17.6 & 17.3 & 1.02 & 26 & -7.1 \\
\hline
\end{tabular}
\end{table*}

Table 1 lists the clusters' basic integrated properties: magnitudes were derived at 12 
arcsec, and (V-I) at the smaller radius of 8 arcsec (but no smaller due to the clumpiness of the 
clusters).
Prior to undertaking the aperture photometry analysis of the clusters, 
 bright stars (i.e. with i' $<20$) were automatically clipped out 
of the images and replaced with a local background estimate using the existing
WFS object catalogues to drive the clipping algorithm.

The INT-WFS data were taken with Harris V (V$^\prime$) 
and Sloan-Gunn i$^\prime$ filters. The colour transformations  
applied to determine \citet{Landolt92} Johnson V and Cousins I magnitudes were
 I = i$^\prime$ - 0.101(V-I) and V = V$^\prime$ - 0.005(V-I).

\begin{figure}
 \includegraphics[angle=-90,width=90mm]{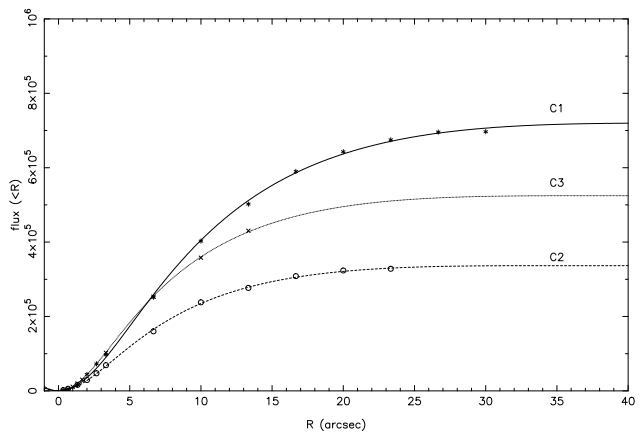}
 \vspace{2pt}
 \caption{King profiles for the clusters where the flux is measured
 simply in ADU counts}\label{Fi:profiles}
\end{figure}

\begin{figure}
 \includegraphics[angle=0,width=80mm]{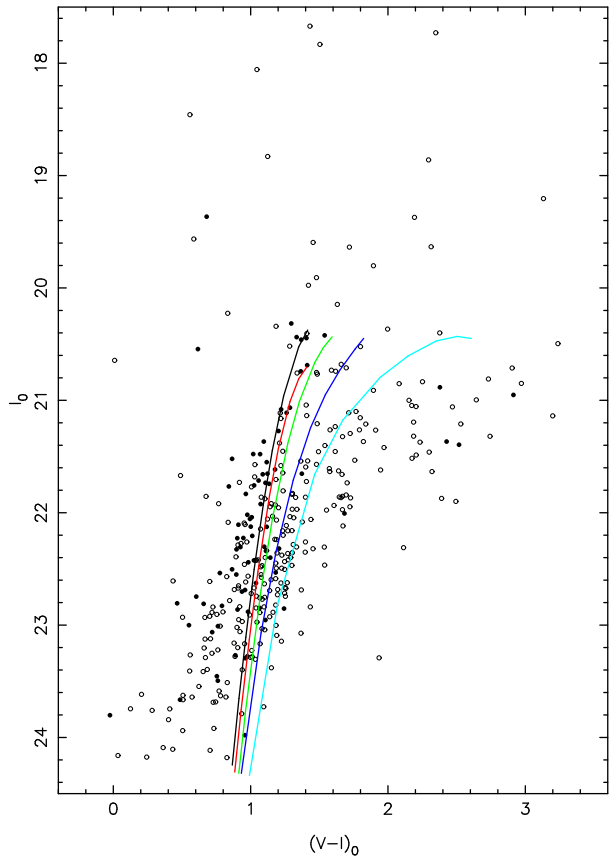}
 \vspace{2pt}
 \caption{A colour-magnitude diagram for C1 (filled points are within
20 arcsec of the cluster centre) and surrounding field (open points
are within a 130 arcsec square region around the cluster).
Photometry was obtained using DAOPHOT (IRAF implementation)
profile fitting, and
has been corrected for foreground extinction
via Schlegel et al. (1998).  Overplotted are isochrones for
galactic globular clusters from \citet{DaCostArmandroff90}
shifted to an M31 distance modulus of $\mu_0=24.47$ and
representing a range of metallicities, from left to right:
[Fe/H]=-2.17 (M15);
-1.91 (NGC 6397); -1.58 (M2); -1.29 (NGC 1851) and $-0.71$ (47 Tuc).
The RGB locus of the C1 is clearly delineated, and consistent
with an old, metal-poor population at the distance of M31.}\label{Fi:cmd1}
\end{figure}

\begin{figure}
 \includegraphics[angle=0,width=80mm]{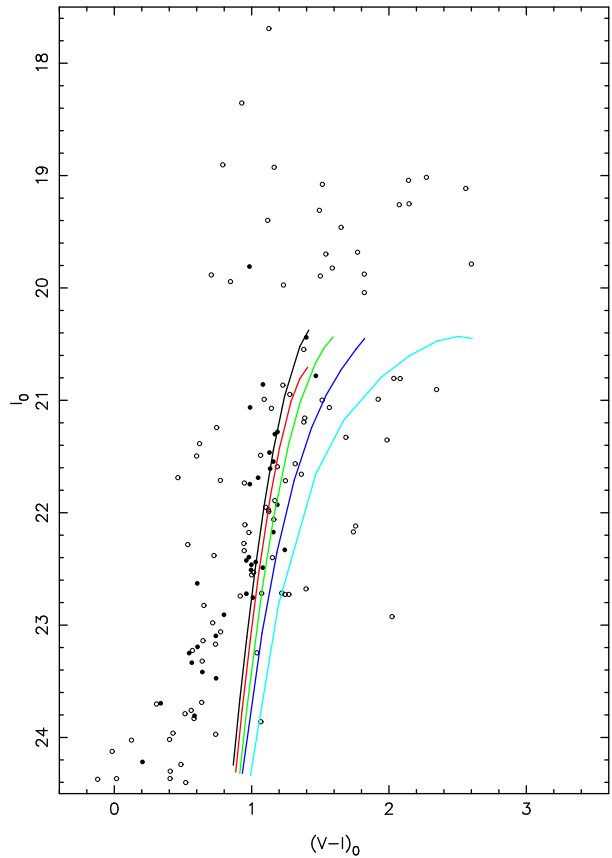}
 \vspace{2pt}
 \caption{CMD for C2, similar to figure 4, but in this case the radius
used to define the cluster region was 14 arcsec.  Although
fewer points, a reasonably clear (metal-poor) sequence is seen.}\label{Fi:cmd2}
\end{figure}

\begin{figure}
 \includegraphics[angle=0,width=80mm]{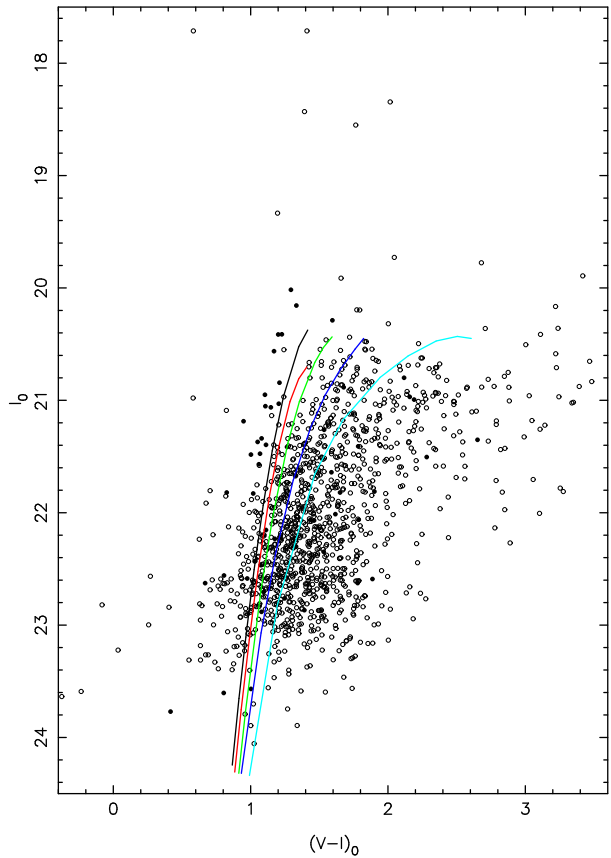}
 \vspace{2pt}
 \caption{CMD for C3, similar to figure 4, again defining the cluster (filled
points) to be within 14 arcsec radius.  In this case the quality
of the image was poorer, and the crowding due to M31 disk/spheroid
stars much worse.  Nonetheless, the very blue (metal-poor) RGB
locus of the cluster is clearly seen.
}\label{Fi:cmd3}
\end{figure}

Empirical King Profiles:
 \[
   \Sigma (r) = \Sigma _{0} \left[  \frac{1}{\left(1 + \left( r/r_{c} \right)^2 \right)^{1/2}} - \frac{1}{\left(1 + \left( r_{t}/r_{c} \right)^2 \right)^{1/2}}       \right]^2
 \]
 
 were fit to the aperture photometry, 
from which ${R_{h}}$ was determined (figure \ref{Fi:profiles}). 
There are some uncertainties associated with the values: 
In the case of C2, likely foreground contamination
from stars was removed first. The cluster C3 lies at the
edge of the INT field but the fractional area missing is small allowing us to obtain
photometry only to a radius of 13 arcsec.

\section{Discussion}

These objects are unusual because, whilst all three clusters have magnitudes 
and colours typical of the M31 GC population, their half-light radii $(R_{h})$
are considerably greater. The data 
available for the MW  (\citealt{vandenbergh96} and references therein) 
shows that there are several GCs (which are also generally at large Galactocentric
radius) with half-light radii 
in excess of 15 pc, but these MW clusters are considerably fainter, and rather more compact,
than the newly found clusters in M31.
The very distant, and extended, Galactic globular
NGC 2419 stands as a notable exception, being considerably more luminous. 
Conversely the newly discovered extreme MW cluster \citep{Willmanetal04} has similar half-light radius
but is some 4 magnitudes fainter.

\begin{table}
 \centering
 \begin{minipage}{84mm}
  \caption{Derived King profile fit parameters for the clusters, giving core radii $r_{c}$, tidal radii
  $r_{t}$, the concentration parameter (c) = log$(r_{t}/r_{c})$, $\Sigma_{0}$ -  the model scale
  surface brightness (obtained by converting counts in 12 arsec radius to solar luminosities, 
  assuming a distance of 780 kpc),  $\rho_{0}$ the central luminosity density, and equivalent
  total integrated V magnitude $M_{V}$ from the profile fit.}
  \begin{tabular}{@{}lllllll@{}}
  \hline
   ID & $r_{c}$ & $r_{t}$ & c & $\Sigma_{0}$ & $\rho_{0}$ & $M_{V}$  \\
      & pc & pc & & L$_{\odot} pc^{-2}$ & L$_{\odot} pc^{-3}$ & \\
 \hline
 M31WFS-EC1   & 23 & 166 & 0.86  & 29 & 0.53 & -7.3 \\
 M31WFS-EC2   & 17 & 132 & 0.89  & 70 & 1.73 & -7.8 \\
 M31WFS-EC3   & 16 & 140 & 0.94  & 42 & 1.12 & -7.2 \\
\hline
\end{tabular}
\end{minipage}
\end{table}

The distinction between different types of stellar cluster is becoming increasingly
complex, with even the boundary between open and globular clusters sometimes being blurred.
Borderline objects have been identified, such as BH 176 \citep{Ortolanietal95} which
could be a true globular cluster or a very old open cluster. The situation is complicated
further by $\omega$ Cen, which although long classified as a GC, has many unusual features, such as
a wide metallicity spread of its component stars\citep{SuntzeffKraft96}, 
a characteristic shared by the bright M31 globular cluster G1 \citep{Meylanetal01}.

More recently, new classes of object have been identified which begin to fill in the gap
in the parameter space between classical GCs and dwarf galaxies.
For example, \citet{Mieskeetal02} measured the properties of ultra-compact objects (UCOs)  
[or ultra-compact dwarfs  - UCDs] in the Fornax cluster, which are as 
bright as dwarf ellipticals but very much more compact. \citet{MartiniHo04} describe 
a population of very massive globular clusters in NGC 5128, whose
properties overlap with those of nucleated dwarfs and UCDs. 
At the fainter end of the luminosity scale
 \citet{BrodieLarsen02} described a new class of star cluster found in 
HST studies of two lenticular galaxies, NGC 1023 and NGC 3384, lying at distances of $\sim$ 10 Mpc. 
 These clusters occupy a region of  parameter space having an 
absolute V magnitude fainter than ${\rm \approx -7}$, a half-light radius greater than 7--15 pc, 
and have been christened ``faint fuzzies" (FFs). 

\begin{figure*}
 \includegraphics[angle=-90,width=150mm]{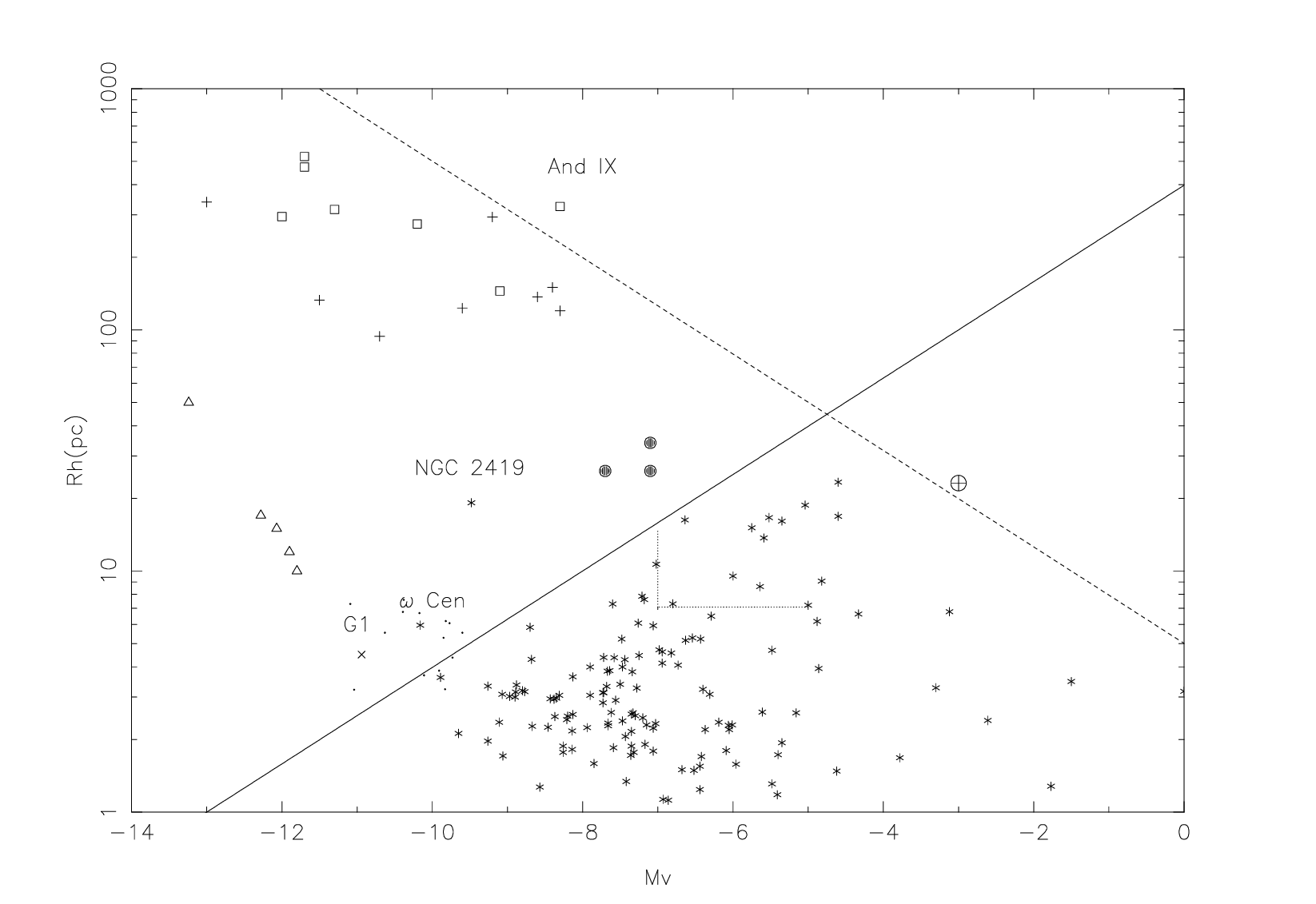}
 \vspace{2pt}
 \caption{Plot of log $R_{h}$ against $M_{V}$ for the new extended clusters (filled circles),
 MW GCs (asterisks, \citealt{vandenbergh96}), MW associated dSphs (plus signs, \citealt{Irwinetal95}), 
 M31 associated dSphs (open squares, \citealt{Caldwelletal92}; \citealt{Caldwell99}),
 the M31 GC G1 (cross signs, \citealt{Larsen01}), UCOs (triangles, \citealt{Mieskeetal02}), 
 the massive GCs in NGC5128 (points, \citealt{MartiniHo04}) and the
 newly discovered MW companion (cross in circle, \citealt{Willmanetal04}). 
 The M31 dSph, AndIX, is also included, using an exponential fit to the
 radial profile determined by us from the INT-WFS  catalogues 
 and using the luminosity from Zucker et al. (2004).
  In addition the line of the equation ${\rm log} R_{h} = 0.2M_{V} + 2.6$
 is plotted (solid line, from \citealt{vandenBerghMackey04}), 
 and that for a value of constant average surface brightness within $R_{h}$
 (dashed line) illustrating one selection effect. The dotted L-shape indicates the region where FFs are found.
 The Galactic GCs $\omega$ Cen and NGC 2419 are also labelled. 
 }\label{Fi:plot}
\end{figure*}

The relationship between all these object types, and MW and M31 dSph galaxies is illustrated
 in Figure \ref{Fi:plot}.
The new extended clusters are shown (solid circles), along with
MW GCs from \citet{vandenbergh96}, dSphs associated with the MW (\citealt{Irwinetal95}), 
the dSphs associated with M31 (\citealt{Caldwelletal92} and \citealt{Caldwell99}), 
and G1 (\citealt{Larsen01}).
The region that would be occupied by the FFs is to to the right of, and above, the dotted
lines on the plot (although they are distinguished from the more diffuse MW GCs in terms
of their numbers, higher metallicities and disky kinematics).

In terms of morphology and luminosity the new 
extended clusters are rather more extended
 and more luminous the FFs (although the magnitudes of the latter may well
 be underestimates by several tenths of a magnitude due to the aperture
 corrections chosen for photometry). 
Furthermore, Brodie and Larsen's FFs are moderately metal rich [Fe/H]$\sim$ -0.6, 
and consequently have red integrated colours, $(V-I)_{0}$ 
of $\approx 1.3$.

Looking more broadly at figure \ref{Fi:plot} we see a variety of objects
in the region between classical GCs and dSphs. A number of origins have
been proposed to explain the intermediate forms.
\citet{Zhao98} has proposed that NGC 2419, the unusual MW GC,
with a half-light radius of $\approx$ 19 pc may have been accreted from a
putative ``Ancient Sagittarius Galaxy", along with some of the fainter GCs:
Terzan 7, Arp 2, Terzan 8 (which are part of the main body of Sgr) and
Pal15.
One might therefore speculate that these new M31 objects formed in
a dwarf galaxy, which may have since merged with M31. For similar
reasons, \citet{BrodieLarsen02} have also suggested that their FFs may have formed
in dwarfs before joining the host galaxies, although they also consider an
alternative theory in which FFs form preferentially in lenticular
galaxies with well-developed disks. However, we should point out that by no means
are all dwarf galaxy GCs unusually extended.

Other possible scenarios to form extended clusters involve either stripping of dwarfs to form smaller 
systems, or the merger of star clusters to form larger systems. 
For example, both G1 and $\omega$ Cen GCs are unique, with suggestions that 
they are the cores of stripped dwarfs 
(\citealt{BekkiFreeman03}; \citealt{BekkiChiba04}; \citealt{IdetaMakino04}; \citealt{Tsuchiyaetal04}).
\citet{MartiniHo04} speculate that due to the presence of
extra-tidal light, that some of their massive GCs in NGC 5128 may be the nuclei of stripped dwarfs.
Numerical simulations by \citet{Bassinoetal94} suggest that the nuclei of dwarf nucleated galaxies survive 
encounters with a giant galaxy, and they propose that such a scenario may explain the large number of
 GCs in many galaxies.
 Depending on the characteristics of the progenitor dwarf, and of the encounter orbit,
a range of remnant clusters can be produced. In their models, the surviving clusters are more massive and have 
larger tidal radii than globular clusters (or our new extended clusters). However, they suggest that 
more concentrated nuclei and closer encounters with the centre of the giant galaxy might produce more compact
clusters, such as GCs. Indeed, it has been proposed that M54 is the nucleus of the Sagittarius 
dwarf (\citealt{BassinoMuzzio95}, but see \citealt{Monacoetal05} for a contrary view).

Such a scenario has important implications. For example, 
it has been suggested that if, as postulated, all galaxies have central
black holes, such stripped objects should have black holes at their cores, and \citep{GebhardtRichHo02} 
report to have found evidence, albeit controversial,  for such a black hole in G1. 
It is also believed that GCs, unlike dSphs, do not possess a 
dark matter halo, leaving open the question of the dark matter content for any 
intermediate forms, such as those listed above. However, \citet{BekkiChiba04} found that any progenitor dwarf 
would have to have had its
DM halo stripped to form a G1-like cluster. 

It has been argued that G1 and $\omega$ Cen can be better explained as the product of merger events.
For example, \citet{Baumgardtetal03} propose a model for G1 involving merger of two smaller GCs 
(and argue that there is no need to invoke a central BH). 
\citet{Fellhauer04} has suggested that $\omega$ Cen could be the product of a merger of
``super star clusters", themselves created in a starburst event triggered by the accretion
of a dwarf satellite, which derives its diverse stellar population by capturing old
field stars from the dwarf.
Furthermore, since the progenitor dwarf itself may have been the product of merging, 
it would retain traces of these  past merging events. \citet{Mieskeetal04} find that their results are
consistent with both stripping and the merged stellar super-clusters scenarios for UCOs.

Does any of this help in understanding our new objects?
The two mechanisms discussed above by which extended clusters may be produced
from other objects (by nurture, as it were), involve either merging of massive star clusters or
stripping of dSphs.
\citet{FellhauerKroupa02}, using N-body simulations, find that they can 
form FF-like objects from the merger of the ``super star clusters" noted above, such as might
have their origins in galaxy-galaxy interaction. 
Formation in such a single interaction event might explain why a population of extended clusters
is found in M31 but not in the MW.
In the Fellhauer and Kroupa picture the FF clusters are subsequently tidally
truncated to obtain the observed cluster sizes, which may not be relevant given
the large galactocentric radii of the M31 clusters.
On the other hand, \citet{vandenBerghMackey04} have noted that only two
Galactic GCs lie above the equation
log $R_{h} = 0.2M_{V} + 2.6$
on the plot in figure \ref{Fi:plot},
$\omega$ Cen and NGC 2419. 
They prefer the stripped dwarf model for these two
objects and suggest that clusters above the line can be best explained by such stripping, such
as that modelled in \citet{Bassinoetal94}.

The question naturally arises as to what distance from M31 could clusters such as ours 
survive against tidal disruption, which obviously relates to the issue of how long-lived 
such clusters can be.
If we take a typical GC mass-to-light ratio of 2, the absolute magnitude values in table 1 give 
masses for the
clusters of $\approx1.5 \times 10^{5}M_{\odot}$. 
Employing a logarithmic halo model and assuming a circular velocity 
of 250 km~s$^{-1}$, we find 
a galactocentric distance of  $\approx 25$ kpc is required
to produce the tidal radii observed.
This is consistent with the radii at which we observe our clusters,
and also the fact that none have been found at smaller distances.
 
The above assumes a $M/L$ typical for GCs, but it is also possible, particularly
if the new clusters are stripped dwarfs, that they could have some dark matter
component.
In this regard it is interesting that
\cite{Willmanetal04} suggest that their 
object may be an extreme dwarf galaxy due to its position on the $M_{V}$/$R_{h}$ plot (their figure 10),
an argument which might also be applied to our new clusters. 

The fact that the similarly extended but much fainter Willman object 
has been discovered in the MW, but no objects similar to our M31 clusters strongly indicates that no such
population exists in our Galaxy.
This suggests that there might
be important differences between the formation and evolution of M31, in comparison with the MW, to allow
for this new cluster population for form and survive. 
This could be established by searching for similar clusters in more distant galaxies covering
a range in Hubble type and environment, but here it becomes increasingly difficult to distinguish
such clusters from background galaxies.

\section{Conclusion}

We have presented three newly discovered globular-like (in terms of luminosity and colour)
clusters in the halo of M31 with unusually large half-light radii, of around 30 pc (compared to typical
GC values between 1 and 7 pc). 
These objects 
start to fill the gap in parameter space between (negligible dark matter) classical GCs 
and (dark matter dominated) dwarf spheroidals.  This is
a region which in recent times has also been encroached by a variety of new objects, such as the
UCOs in Fornax and massive GCs in NGC 5128. 
Our clusters are lower luminosity systems than these, and are
rather brighter and more extended than the so-called faint fuzzy clusters  
(FFs), found by  \citep{BrodieLarsen02} in two lenticular galaxies, 
and also bluer indicating a lower metallicity.

The extended M31 clusters have no known analogues in the Milky Way, where such clusters 
would certainly have been discovered if they existed,
unless hidden by the plane of the Galaxy.  
This suggests that they could hold important clues to the differing formation
histories of these galaxies.
If these clusters were not born with their present morphology then
one may speculate that they are the stripped cores of cannibalised dwarf 
spheroidal galaxies, or the products of cluster mergers perhaps themselves created
in a previous interaction of a gas-rich companion with M31.

Unlike the UCOs and faint fuzzies, these objects are comparatively nearby in M31,
and hence (like G1) open to detailed, resolved, investigation by both HST and ground-based instruments.
Further, ongoing, observations will reveal more about the nature of these intriguing objects:
metallicity, HB morphology, systemic and internal kinematics.

\section*{Acknowledgments}

AH is funded by a Particle Physics and Astronomy Research Council (PPARC) studentship. 
The research of AMNF has been supported by a Marie Curie Fellowship
of the European Community under contract number HPMF-CT-2002-01758.

This research has made use of the NASA/IPAC Extragalactic Database (NED) 
which is operated by the Jet Propulsion Laboratory, California Institute of 
Technology, under contract with the National Aeronautics and Space Administration.

The Isaac Newton Telescope is operated on the island of La Palma by the Isaac Newton Group in 
the Spanish Observatorio del Roque de los Muchachos of the Instituto de Astrofisica de Canarias 

We would also like to thank the referee, Sydney van den Bergh, whose comments greatly improved the quality of this paper.

\end{document}